\documentclass[pre,aps,floats,twocolumn,floatfix]{revtex4-1}
\usepackage{graphicx,amssymb,amsmath,ifthen,rotating}
\usepackage[active]{srcltx}

\begin{document}

\title{D-dimensional self-gravitating lattice gas in general relativity}

\author{Benaoumeur Bakhti}
\email{benaoumeur.bakhti@univ-mascara.dz, \\ bbakhti@uni-osnabrueck.de}
\affiliation{G2E Lab, FSNV and Department of Physics, University of Mustapha Stambouli, Mascara 29000, Algeria}

%

\begin{abstract}
Using a lattice equation of state combined with the D-dimensional Tolman-Oppenheimer-Volkoff equation and the Friedmann equations, we investigate the possibility of the formation of compact objects as well as the time evolution of the scale factor and the density profile of a self-gravitating material cluster. The numerical results show that in a $2+1$ dimensional spacetime, the mass is independent of the central pressure. Hence, the formation of only compact objects with a finite constant mass similar to the white dwarf is possible. However, in a $3+1$ dimensional spacetime, self-gravity leads to the formation of compact objects with a large gap of mass and the corresponding phase diagram has the same structure as the one for Neutron Star. The results also show that beyond certain critical central pressure, the star is unstable against gravitational collapse, and it may end in a black hole. Analysis of spacetimes of higher dimensions shows that gravity has the stronger effect in $3+1$ dimensions. Numerical solutions of the Friedmann equations show that the effect of the curvature of spacetime increases with increasing temperature, but decreases with increasing dimensionality beyond $D=3$.
%
\end{abstract}
\maketitle
\section{Introduction}
The theory of critical phenomena provides powerful tools for understanding many phenomena in astrophysics and cosmology \cite{Vega/etal:2001,Unruh/Schutzhold:2007}. One example that has attracted great interest is the formation of compact stellar objects from critical collapse. Since the seminal work of Choptiuk \cite{Choptuik:1993} which has connected the gravitational collapse to the theory of phase transition, critical phenomena such as universality and scaling have been proved for many cosmological processes and in particular the black hole solutions of the quantum field theory coupled to the Einstein equations of gravity. But, despite being widely studied, many questions remain to be elucidated, and in particular how does a short distance regularization of the gravitational potential affect the mass threshold of formation of compact objects in cosmology. 

In the theory of general relativity \cite{Weinberg:2008,Weinberg:1972}, the evolution of the Universe and the formation of large-scale structures are described by the Friedmann equations \citep{Friedman:1922} and the Tolman-Oppenheimer-Volkoff equation \cite{Tolman:1939,Oppenheimer/Volkoff:1939}, respectively. However, for these equations to be solved, they need to be supplemented by a thermodynamic equation of state (EOS). It is well known that many of the physics of black holes and other astrophysical objects in the early and late Universe can be understood by simply relying on the law of classical thermodynamics \cite{Wald:1994, Lust/Vleeshouwers:2019,Chavanis:2006}. In fact, it has been shown that the formation of many astrophysical objects is relying on a polytropic EOS in which the pressure depends upon the density in the simple form $p=K\rho^{\gamma}c^2$ where $K$ and $\gamma$ are constants. The early Universe is assumed to be described by a linear EOS $p=w\rho c^2$ where the parameter $\omega$ takes the values $\omega=1/3$ for radiation, $\omega=0$ for no-relativistic matter, and $\omega=-1$ for dark energy. The formation of compact objects is a consequence of gravitational collapse which is fully driven by gravitational attraction. But if the gravity is not too high, the collapse is stopped midway by the quantum mechanical forces due to the Pauli exclusion principle, which plays a dominant role at short distances.  

Due to its singularity at a short distance and long-range, the gravitational interaction
exhibits complex phenomena such as the possibility of the existence of two different phase transitions (with two different critical points) in the canonical (closed system) and microcanonical ensemble (isolated system) \cite{Chavanis:2006}. In addition, a system in the microcanonical ensemble may exhibit a negative specific heat \cite{LyndenBell/Wood:1968,Thirringn:1970}, leading to the formation of exotic astrophysical objects such as black dwarfs. In this letter, we evade the singularity of the gravitational interaction by introducing a short distance regularization that mimics the Pauli exclusion principle. The short distance regularization is introduced by considering a lattice gas description \cite{Chavanis:2014b, Bakhti/etal:2018,Bakhti/etal:2015c,Bakhti/etal:2013a,Bakhti/etal:2012} of the system instead of classical gas of point particles. Besides that it has not been considered previously in a fully relativistic setup, the lattice EOS has many advantages including that in the non-relativistic limit, it is fully consistent with the known results of gravitational collapse \cite{Chavanis:2014b,Bakhti/etal:2018} and at low-density, it produces the Lane-Emden equation for polytropic fluids \cite{Chavanis:2014a,Bakhti/etal:2018}. In addition, in the non-relativistic limit, the results of the lattice EOS \cite{Chavanis:2014b,Bakhti/etal:2018} are in very good agreement with known results of gravitational collapse of quantum systems \cite{Ingrosso/Ruffini:1988}. Our aim in this paper is two folds: first, we want to see how such a regularization affects the formation threshold and the mass range of steller compact objects and black holes in a $D$-dimensional space.  Second, because in the low-density limit, the lattice EOS reduces to the one used for describing the early Universe and the  compact object mass as expected is in the lower range, then the lattice EOS can be used in the fully quantum mechanical description of primordial black hole including its connection to dark matter \cite{Ahrich/etal:2019,Ahrich/etal:2018,Ahrich/etal:2016,Harada/Jhingan:2016,Abada/Nasri:2013,Babu/etal:2007,Calmet/etal:2014,Vega/etal:2001}. This is currently work under progress for future publication. The rest of the paper is organized as follows: In Sec.~(\ref{sec:mod}) and (\ref{sec:eq_stat}), we present the model and the lattice EOS,  respectively. In Sec.~(\ref{sec:tov_eq}), we solve the TOV equation in $2+1 $ and $3+1$ dimensions, and then we generalize the results to higher dimensions. In Sec.~(\ref{sec:fr_eq}), we present a solutions of the Friedmann equations in $2+1 $ and $3+1$ dimensions with different  curvatures of the spacetime. A brief discussion of higher dimensional systems will be also presented. Finally, we conclude with a summary and an outlook for future work.

\section{Model}\label{sec:mod}
Our system consists of a material cluster of $N$ particles confined into a spherical volume of radius $R_c$. 
The total volume is subdivided into $n$ cells each with elementary volume $V_c$. Each cell can be either occupied by a particle of mass $m$ or empty. The particles are subject to short-range hard core exclusion interaction and a long-range gravitational attractive interaction. In $D$-dimensional space, the gravitational potential at distance $r$ from the origin is given by
\begin{align}
g(r) =-\frac{d\mathcal{U}(r)}{dr}= -\frac{G_DM(r)}{r^{D-1}}.
\end{align}
and it is generated by the mass $M(r)$ enclosed in the sphere of radius $r$
\begin{align}\label{eq:mass_r}
M(r)=\rho_0\!\int_{0}^{r}dr'S_{\!D}r'^{D-1}\rho(r'),
\end{align}
where $\rho_0=m/V_c$ and
\begin{align}
S_{\!D}=\frac{2\pi^{D/2}}{\Gamma(D/2)}.
\end{align}
is the surface of the unit hypersphere in $D$ dimensions, $\Gamma$ is the Euler gamma function and $G_D$ is the gravitational constant. The total mass of the system is
\begin{align}\label{eq:total_mass}
M_T=\int_{0}^{R_c}dr'S_{\!D}r'^{D-1}\rho(r').
\end{align}
For numerical purposes, it is more convenient to write Eq.~(\ref{eq:mass_r}) in the form
\begin{align}\label{eq:mass_r1}
\frac{dM(r)}{dr}=\rho_0 S_{\!D}r^{D-1}\rho(r).
\end{align}
and use the following scaled quantities
\begin{align}\label{eq:tov_scaling}
\bar{M}=\frac{M}{M_\odot},\hspace{5mm}\bar{r}=\frac{r}{R_0}, \hspace{5mm}\bar{\epsilon}=\frac{\epsilon}{\epsilon_0}=\frac{\rho c^2}{\epsilon_0}.
\end{align}
where $M_\odot = 1.989\hspace{1mm} 10^{30}\hspace{1mm} kg$ is the mass of the sun, $R_0=GM_\odot/c^2=1.47km$ is one half the Schwartzschild radius of the sun, $\epsilon=\rho c^2$ is the energy density and $\epsilon_0$ has a dimension of energy density and it has been introduced also for a numerical purpose. With this scaling, Eq.~(\ref{eq:mass_r1}) becomes
\begin{align}\label{eq:mass_r2}
\frac{d\bar{M}(\bar{r})}{d\bar{r}}=\alpha_D\bar{r}^{D-1}\bar{\epsilon},
\end{align}
where $\alpha$ is a constant and it is given by
\begin{align}
\alpha_D=\frac{\epsilon_0S_D R_0^{D}\rho_0}{M_\odot c^2}.
\end{align}
\section{Equation of state}\label{sec:eq_stat}
The Thermal equilibrium of any physical system at finite uniform temperature is described by an EOS that relates the pressure to the density. Many astrophysical systems (referred to as polytropic fluids) are described by polytropic equations of state of the form
\begin{align}\label{eq:poly_eq_stat}
p=K\rho^{\gamma}c^2
\end{align}
where $p$ is the pressure, $\rho$ is the density, $c$ is the speed of light and $K$ is a constant of proportionality.
Upon varying the polytropic index $\gamma$, Eq.~(\ref{eq:poly_eq_stat}) describes many astrophysical objects such as white dwarf ($\gamma=5/3$ for the non-relativistic case and $\gamma=4/3$ for the relativistic one) and Neutron star ($\gamma=5/3$ for non-relativistic EOS and $\gamma=1$ for a relativistic EOS) \cite{Silbar/Reddy:2003,Shapiro/Teukolsky:2004}.
The early stage of the Universe is described by a linear EOS
\begin{align}\label{eq:poly_eq_stat1}
p = w\rho c^2.
\end{align}
The constant $w$ lies in the range $-1\leq w \leq 1$. The constant $w$ takes the values $w=-1, 0$ and $1/3$ for a Universe dominated by dark energy, non-relativistic matter, and radiation, respectively but the value of $w$ which corresponds to the threshold of black hole formation is still debatable. Depends on the values of $w$, the Universe evolves from an inflation era dominated by dark energy to matter-dominated era passing by a radiation-dominated era (principally photons and neutrinos).\\
In our analysis, we will use instead a lattice description
in which the minimum distance between the particles is constrained to a short distance scale $\sim V_c^{1/D}$ \cite{Bakhti/etal:2018}. $V_c$ can be considered as the effective volume of the particle. The system is described in any dimension by the universal EOS \cite{Chavanis:2014a, Chavanis:2014b,Bakhti/etal:2018}
\begin{align}\label{eq:eq_state}
p=-\frac{k_BT}{m}\rho_0\ln\left(1-\frac{\rho}{\rho_0}\right).
\end{align}
where $k_B$ is the Boltzmann constant and $T$ is the temperature. It has been shown in previous studies \cite{Chavanis:2014b,Bakhti/etal:2018}, that the low-density limit of Eq.~(\ref{eq:eq_state}) produces results that fit very well with the Lane-Emden equation that describes polytropic fluids in Newtonian hydrostatic equilibrium.\\
It is also convenient here to write Eq.~(\ref{eq:eq_state}) in a scaled form
\begin{align}\label{eq:scaled_eq_stat}
\bar{p}=-\bar{T}\ln(1-\bar{\epsilon}),
\end{align}
where
\begin{align}\label{eq:scaled_EOS}
\bar{p}=\frac{p}{\epsilon_0},\hspace{5mm}\bar{\epsilon}=\frac{\rho c^2}{\epsilon_0},\hspace{5mm}\bar{T}=\frac{k_BT}{mc^2},
\end{align}
and we set
\begin{align}\label{eq:eps0_rho0}
\epsilon_0=\rho_0 c^2.
\end{align}
In the low-density limit, Eq.~(\ref{eq:scaled_eq_stat}) reduces to the EOS of classical gas of point particles which has the same form as Eq.~(\ref{eq:poly_eq_stat1}), with temperature to be identified with the constant $\omega$. The latter equation has been used to describe the early Universe and it is also known to provide a very good approximation in the relativistic description of Neutron stars.\\
In the continuum EOS, the particles are classical point particles, so during the collapse, all the cluster collapses to one point and the mass ends up in a delta function profile. In the lattice EOS, the separation between any two particles is constrained to some minimum distance (constraint in position space). This classical lattice description mimics the quantum mechanical exclusion principle operating for the fermionic matter. A detailed studies of the corresponding Fermi gas at nonzero temperatures have been worked out in \cite{Chavanis:2002,Chavanis:2004} in the Newtonian limit and in \cite{Chavanis:2020,Alberti/Chavanis:2020a,Albertini/Chavanis:2020b,Chavanis/Alberti:2020c} in general relativity. The lattice description evades the singularity in the gravitational potential which leads to the total gravitational collapse. Our  goals with using a lattice EOS is first to see whether the lattice EOS stops midway gravitational collapse so the cluster ends in a compact object with finite mass and size (as did the quantum mechanical forces),  and our second goal is to see how does such a regularization affect the formation threshold and the mass range of compact objects and black holes.

\section{TOV Equations}\label{sec:tov_eq}
In a general relativistic description, the hydrostatic equilibrium of a relativistic fluid is described by the TOV equation \cite{Tolman:1939,Oppenheimer/Volkoff:1939}. 
A spherically symmetric self-gravitating perfect fluid in $D+1$-dimensional spacetime is described by the Einstein field equations
\begin{align}\label{eq:einstein_eq}
R_{\mu\nu}-\frac{1}{D-1}R_sg_{\mu\nu}=8\pi G_DT_{\mu\nu},
\end{align}
where $T_{\mu\nu}$ is the $D+1$-dimensional stress-energy tensor and it depends only on pressure and density of the perfect fluid
\begin{align}\label{eq:stress_energy_tensor}
T_{\mu}^{\nu}(r)=diag(-\rho,p,\ldots,p).
\end{align}
$R_{\mu\nu}$ is the Ricci curvature and $R_s$ is the Ricci scalar. The geometry is described by the spherically symmetric metric
\begin{align}\label{eq:metric}
g_{\mu\nu}=-e^{\psi(r)}dt^2+e^{\lambda(r)}dr^2+r^2d\Omega_{D-1},
\end{align}
$\psi(r)$ and $\lambda(r)$ are functions to be determined and $d\Omega_{D-1}$ denotes the canonical metric of the unit hypersphere $\mathsf{S}^{D-1}$ in $\mathbb{R}^D$.
Inserting the stress-energy tensor Eq.~(\ref{eq:stress_energy_tensor}) and the metric Eq.~(\ref{eq:metric}) into the Einstein's equations Eq.~(\ref{eq:einstein_eq}), we get a set of equations that can be solved analytically to get an expressions for $\lambda(r)$ and $\psi(r)$. They are given by \cite{Leon/Cruz:2000}
\begin{align}\label{eq:lambda}
e^{-\lambda(r)}=1-\frac{2G_DM}{r^{D-2}}
\end{align}
where the mass $M$ is given by Eq.~(\ref{eq:mass_r}), and 
\begin{align}\label{eq:psi}
\frac{d\psi(r)}{dr}\!=\!
\frac{2G_DM}{r^{D-1}}\!\!\left(\!D-2+\frac{8\pi r^{D} p}{c^2(D-1)M}\!\right)\!\!\left(\!1-\frac{2G_DM}{c^2r^{D-2}}\!\right)^{-1}.
\end{align}
From the conservation of stress-energy tensor $\nabla_{\mu}T^{\mu\nu}=0$, we get
\begin{align}
\frac{dp}{dr}=-\frac{\rho}{2}\left(1+\frac{p}{\rho c^2}\right)\frac{d\psi}{dr},
\end{align}
which combined with Eq.~(\ref{eq:psi}), leads to the TOV equation
in a $D+1$-dimensional spacetime \cite{Leon/Cruz:2000}
\begin{align}\label{eq:d_tov}
p'=-&\frac{G_D}{r^{D-1}}M\rho\left(1+\frac{p}{c^2\rho}\right)\\
&\times\left(D-2+\frac{8\pi r^{D}p}{c^2(D-1)M}\right)\!\!\left(1-\frac{2G_DM}{c^2r^{D-2}}\right)^{-1}\nonumber.
\end{align}
At low-density and for $c\rightarrow\infty$, Eq. (\ref{eq:d_tov}) reduces to the Newtonian hydrostatic equilibrium condition which in three-dimensions reads \cite{Chavanis:2014b,Bakhti/etal:2018}
\begin{align}
&p'=-G_D\frac{M\rho}{r^{D-1}}.\label{eq:hydr_eq}
\end{align}
Using the scaling Eq.~(\ref{eq:tov_scaling}), the $D$-dimensional TOV equation can be written as
\begin{align}\label{eq:d_tov_scaled}
\bar{p}'=\frac{1}{R_0^{D-3}}&\frac{\bar{M}}{\bar{r}^{D-1}}(\bar{\epsilon}+\bar{p})\left(D-2+\bar{\alpha}_D\frac{\bar{r}^{D-1}\bar{p}}{\bar{M}}\right)\nonumber\\
&\times\!\left(1-\frac{2}{R_0^{D-3}}\frac{\bar{M}}{\bar{r}^{D-1}}\right)^{-1},
\end{align}
where 
\begin{align}
\bar{\alpha}_D=\frac{8\pi\alpha_D}{\rho_0S_D(D-1)}.
\end{align}
To get density (or energy density) and pressure profiles, Eq.~(\ref{eq:d_tov_scaled}) needs to be supplemented by the EOS (\ref{eq:scaled_eq_stat}) and solved numerically out from the origin at $r=0$ to the point $R$ where the pressure falls to zero. The boundary conditions required to solve the two coupled non-linear differential equations Eqs.~(\ref{eq:d_tov_scaled}) and (\ref{eq:scaled_eq_stat}) are
\begin{align}
p(R) = 0,\hspace{10mm} p(0)=p_0,
\end{align}
for the pressure and
\begin{align}
M(R)=M_T=Nm_c,\hspace{10mm}M(0)=0.
\end{align}
for the mass. We can also use boundary conditions for mass or energy density
\begin{align}
\epsilon(0)=\epsilon_0,\hspace{10mm}\rho(0)=\rho_0,
\end{align}
hold at the origin of the material cluster. At the surface of the cluster, boundary condition on $\rho$ (or $\epsilon$) is encoded in Eq.~(\ref{eq:total_mass}).

To solve the TOV equation, we have used the Runge-Kutta (RK45) method. For boundary conditions, we have used the values of the pressure and mass at the origin of the cluster ($p(0)$ and $M(0)$). The choice of $p_0$ can be arbitrary, but only a finite range of $p_0$ leads to physically acceptable solutions. We can also impose boundary conditions on the derivative of the pressure and the mass and they lead to the same solutions \cite{Bakhti/etal:2018}. So, the solution is independent of chosen boundary conditions.  However, the numerical solutions of the TOV equation depend on the central density (that fixes the parameter $\bar{\alpha}_D$) or the central pressure $p_0$. Their values are fixed either by supplementing the TOV equation and EOS with the equation of conservation Eq.~(\ref{eq:total_mass}) (or the corresponding equation for pressure) or by fitting to existing experimental or simulation results.

\subsection{2+1 Dimensions}
\begin{figure}[htb]
\centering
\includegraphics[scale=0.27]{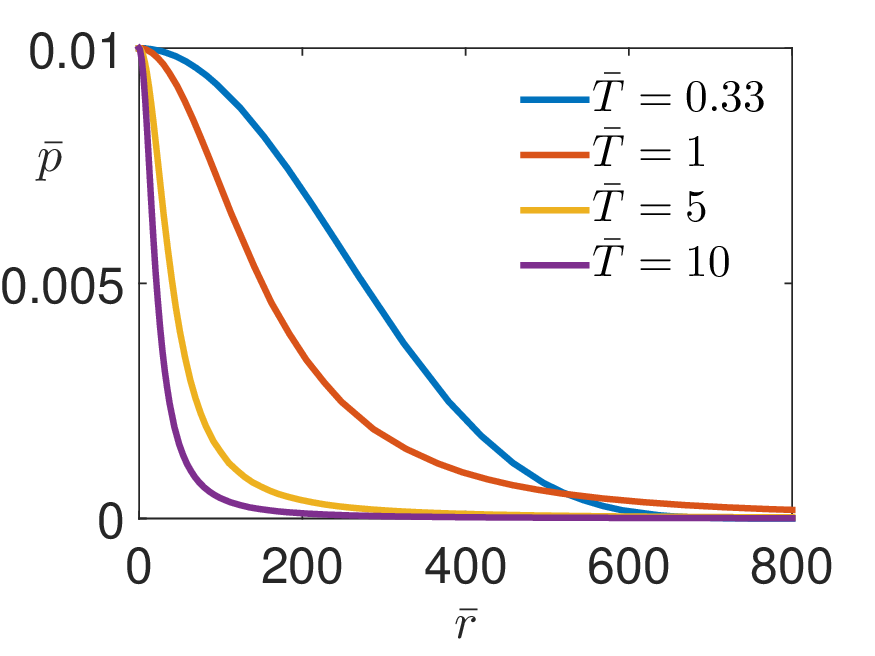}
\includegraphics[scale=0.27]{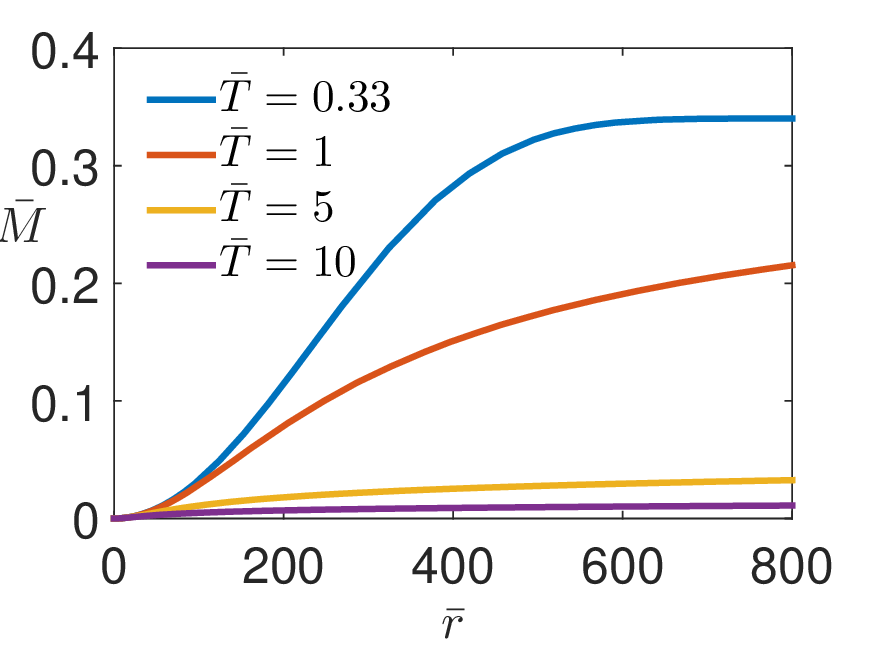}\\
\includegraphics[scale=0.27]{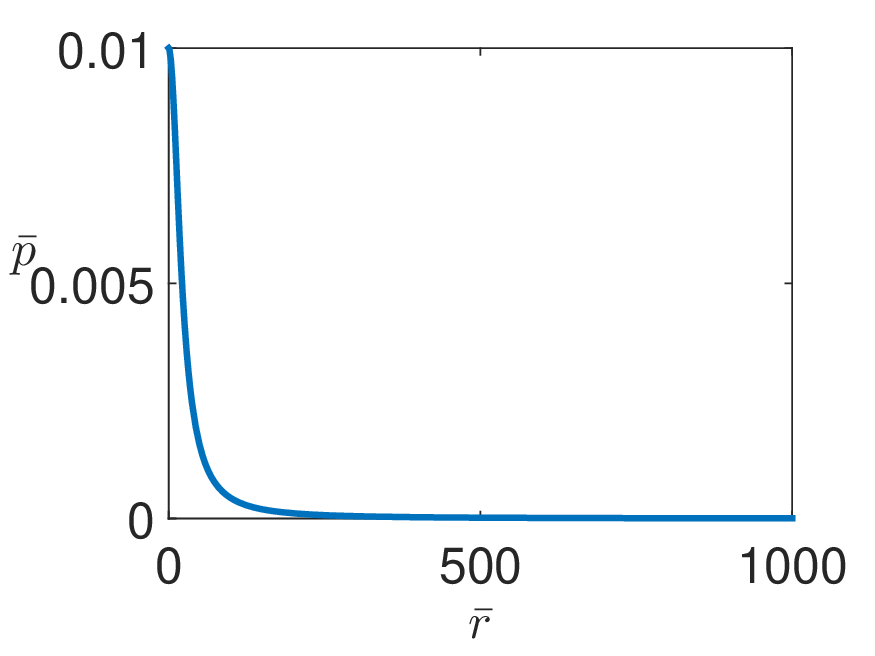}
\includegraphics[scale=0.27]{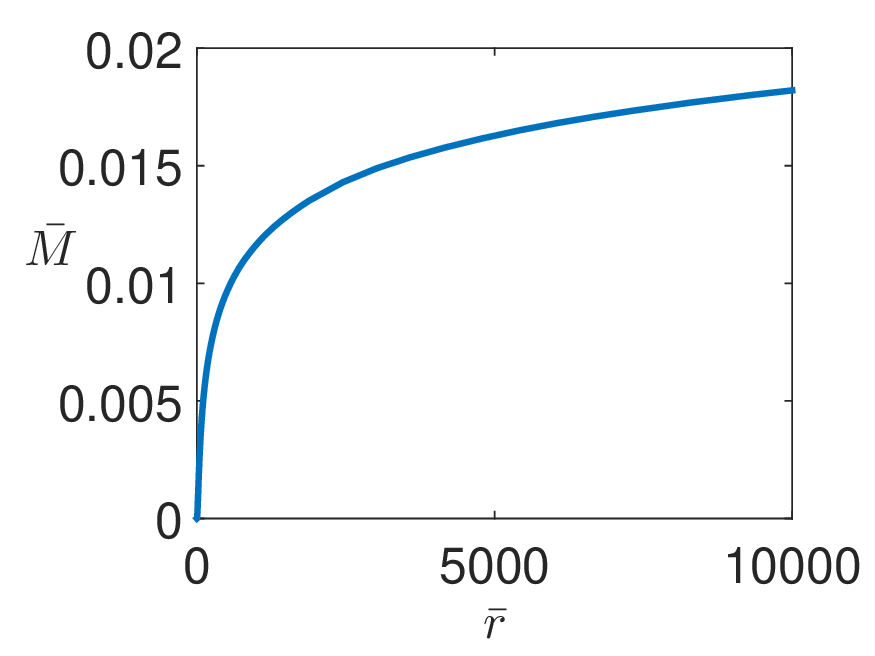}
\caption{\label{fig:fig1} Pressure $\bar{p}$ and mass $\bar{M}=M/M_\odot$ versus scaled distance $\bar{r}=r/R_0$ at different temperatures for lattice system in $2+1$ dimensions. The two panels at the bottom represent pressure and mass of the cluster at very high temperature ($\bar{T}=10$).}
\end{figure}

The $2+1$ dimensional system is a toy mathematical model but it has great physical interest mainly in string theory and quantum gravity. If the spacetime is created by an infinite cosmic string, the $3+1$ dimensional spacetime is reduced to an effective $2+1$ dimensional spacetime \cite{Carlip:2005}. In addition, as pointed out by Jackiw \emph{et al} \cite{Deser/etal:1984,Jackiw:1985}, systems in a hot phase are phenomenologically described by $2+1$ dimensional system. Also, the model could be relevant for the description of large one-dimensional structures that seem to be observed in the Universe, such as strings and vortices, whose interactions are governed by 2+1 gravity \cite{Deser/etal:1984,Jackiw:1985}.\\
Besides that, there are two other good reasons for the interest in the $2+1$ dimensional systems. On one hand, the model admits a black hole solution in an anti-de Sitter space \cite{Banados/etal:1992,Banados/etal:1993,Carlip:1995,minneborg/etal:1998} and on the other hand, it is known to have a simple mathematical structure \cite{Carlip:2005}. These revive the hope to quantize gravity in $2+1$ dimensions \cite{Witten:1988}. And because the model shares some important conceptual features of general relativity in $3+1$ dimensions \cite{Carlip:2005}, then it is believed that quantizing gravity in $2+1$ dimensions will shed some light on the quantum gravity of real systems in $3+1$ dimensions.\\
It is well known that for an asymptotically flat spacetime, there is no black hole solution for the Einstein equations with a polytropic EOS. The aim of this section is to see whether this is true or not for a lattice system. In a three-dimensional spacetime ($D=2$), the TOV equation is reduced to 
\begin{align}\label{eq:2_1_tov}
\bar{p}'=-4\alpha_2c^2R_0\bar{r}\bar{p}(\bar{\epsilon}+\bar{p})(1-2R_0\bar{M})^{-1}.
\end{align}
where the pressure derivative can be inferred from Eq.~(\ref{eq:scaled_eq_stat}) and it is given by
\begin{align}
\bar{p}'=\frac{\bar{T}\bar{\epsilon}'}{1-\bar{\epsilon}}.
\end{align}
The EOS (\ref{eq:scaled_eq_stat}) and Eq.~(\ref{eq:2_1_tov}) constitute a complete set of  coupled non-linear differential equations that can be solved numerically using an appropriate boundary conditions. Numerical solutions are shown in Fig.~(\ref{fig:fig1}), which represents variations of the pressure $\bar{p}$ and mass $\bar{M}$ of the cluster at different temperatures.
We see that for all temperatures, the mass spreads out in space and it reaches a constant value beyond a certain distance from the origin. The pressure drops from its value at the origin (taken here to be $\bar{p}_0=0.01$) to zero at the surface of the massive cluster, exactly where the mass becomes constant. The distance where the mass becomes constant and the pressure is null corresponds to the physical value of the radius $R$ of the compact object. The mass of the compact object is simply the mass enclosed within the sphere of radius $R$. Now, if we take a different value for the central pressure, we get exactly the same value of the mass ($\bar{M}=0.34$) but with a small decrease in the radius $R$, so the density increases and the compact object becomes more compact (denser). The $D=2$ lattice model allows the formation of compact objects. Different temperatures correspond to compact objects with different radii, however, all with the same total mass. In our lattice model, the mass approximately equals $\bar{M}=0.34$. The compact object we have found has the same properties and falls within the mass range of a white dwarf \cite{Shapiro/Teukolsky:2004}. The estimated mass of white  dwarf is in the range $\bar{M}=0.17$ to $\bar{M}=1.33$ with peak at around $\bar{M}=0.6$. The case of high temperature ($\bar{T}=10$) is shown separately in the bottom of Fig.~(\ref{fig:fig1}) for larger values of the scaled distance. As can be seen, at very high temperatures, the mass is dispersed to infinity and never reaches a constant value that corresponds to some compact object. The pressure drops rapidly reaching very small values but never goes to zero. So, at very high temperatures, there is no sign of the formation of black hole neither for any compact object. This can be understood from the fact that at high temperatures, the particles have enough energies to escape from the gravitational attraction of each other and they dispersed to infinity. At high temperatures, the variation of the mass versus the scaled distance crosses between two regimes. First, the mass increases rapidly near the center of the cluster, then increases smoothly beyond a certain distance. The cluster thus has a form of "Core-Halo" structure, with a high-density core surrounded by gas with a smooth density that extends to infinity. Varying the central pressure produces similar curves.\\
For a black hole to be formed, there must be a critical value of the central pressure beyond which the compact object cannot support the increasing gravitational attractions generated by the large mass so it collapses to form black hole. In $2+1$ dimensions, the numerical solution confirms that whatever the value of the central pressure, one gets always (depends on the model) the same value of the mass with the corresponding radius. Thus, the formation of black hole is not possible in the $2+1$ dimensional spacetime of GR. The situation becomes different if we consider an anti-de Sitter space of $2+1$ dimensions, a problem that is now under consideration. In this case, the model has great physical relevance in which the formation of black hole is possible and interesting results have been worked out along this line.
\subsection{3+1 Dimensions}
In the $3+1$ dimensional spacetime of general relativity, the TOV equation becomes
\begin{align}\label{eq:de_density}
\bar{p}'=-\frac{\bar{M}}{\bar{r}^2}(\bar{\epsilon}+\bar{p})\left(1+\alpha_3c^2\frac{\bar{r}^3\bar{p}}{\bar{M}}\right)\left(1-2\frac{\bar{M}}{\bar{r}}\right)^{\!\!-1}.
\end{align}
which can be combined with the EOS to get profiles of mass density, energy density and pressure. 
\begin{figure}[htb]
\centering
\includegraphics[scale=0.27]{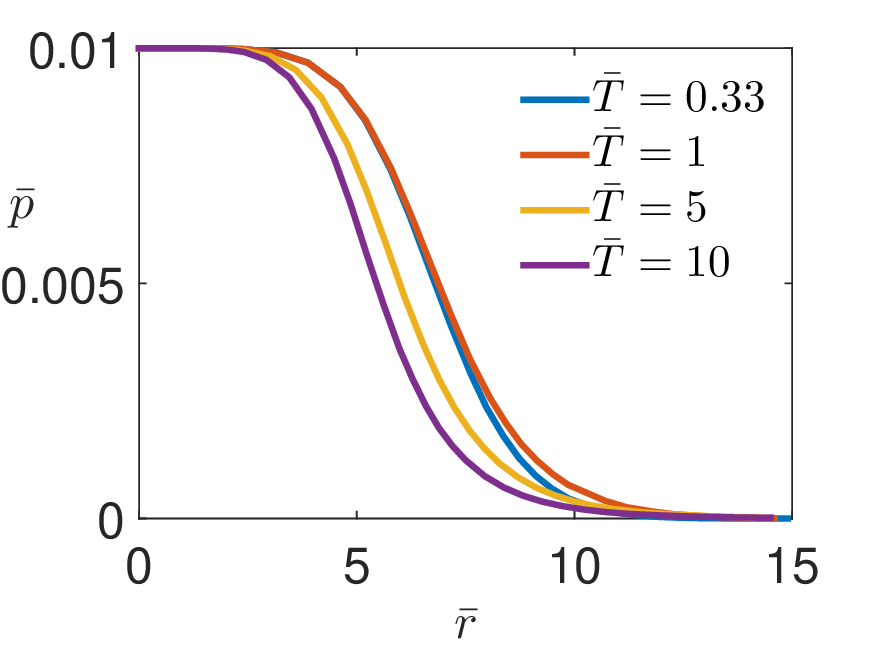}
\includegraphics[scale=0.27]{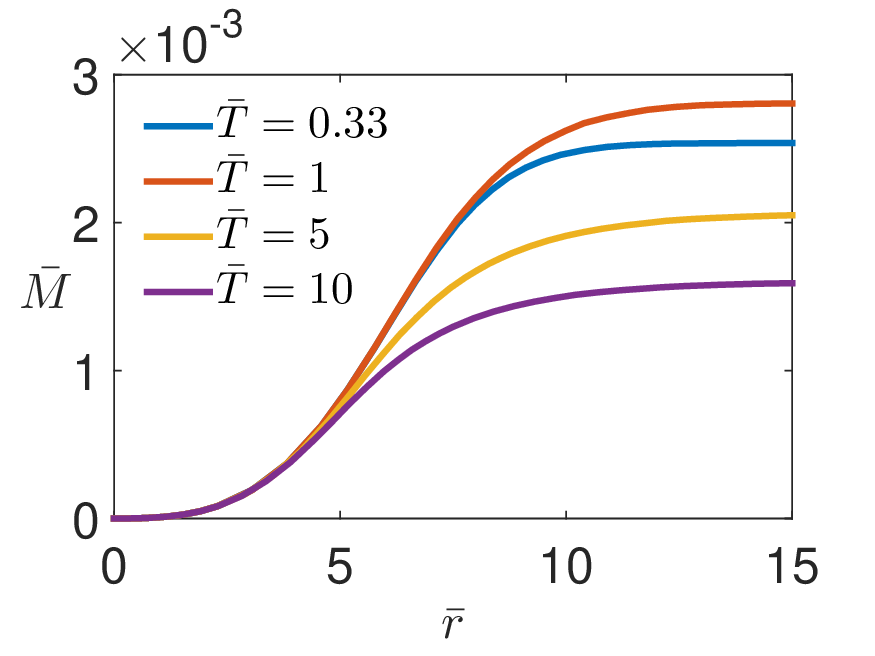}
\caption{\label{fig:fig2} Pressure $\bar{p}$ and mass $\bar{M}=M/M_\odot$ versus scaled distance $\bar{r}=r/R_0$ at different temperatures for self-gravitating lattice system in $3+1$ dimensions.}
\end{figure}
Numerical solutions are depicted in Fig.~(\ref{fig:fig2}). We see that at different temperatures, the masses are reaching constant values almost at the same radius and hence these objects should have different mass densities. If we change the temperature and/or central pressure, we get a different value of the mass density. Contrary to the case of $2+1$ dimensions, the mass of the compact object in $3+1$ dimensions depends on the central pressure and also on temperature.  With increasing the central pressure, the cluster can support more mass. More mass means more gravitational attraction and hence the radius of the cluster decreases. The numerical values point to the conclusion that the resulting compact objects have similar properties to those of Neutron stars and they have the same radii, but they have lower mass compared to Neutron stars.\\
The lattice EOS  $\bar{p}=-\bar{T}\ln(1-\bar{\epsilon})$ allows formation of compact objects with different mass densities that are similar to Neutron stars. In the lattice EOS, the quantum mechanical effects (like Neutron or Electron degeneracy pressure) are hidden in the exclusion effects. The low-density limit of the lattice EOS produces all results of the polytropic EOS \cite{Chavanis:2014b,Bakhti/etal:2018}. The corresponding phase diagram of the compact objects is shown in Fig.~(\ref{fig:fig3}) for different values of $\bar{T}$ and it has qualitatively the same structure as the one for Neutron stars \cite{Parente/Roychowdhury:2011}.\\
\begin{figure}[t]
\centering
\includegraphics[scale=0.4]{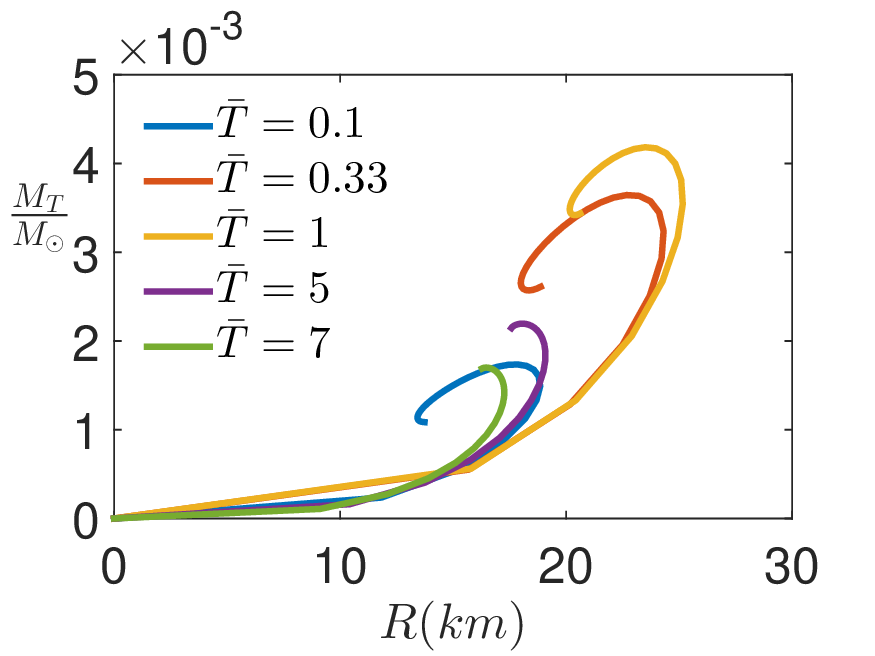}
\caption{\label{fig:fig3} Mass versus radius of the compact objects in $3+1$ dimensions at different temperatures of $\bar{T}$.}
\end{figure}
Limits of stabilities of the self-gravitating cluster at different values of the scaled temperature are shown in Tab.~(\ref{tab:stability_limits}). Beyond these limiting values, the compact objects become in non-equilibrium unstable states and they start collapsing. The radii presented in the table are the values at which the compact objects become unstable. The numerical analysis confirms that a singularity is developed in the mass density profile at the end of the collapse. To test black hole formation in a gravitational collapse in
spherical symmetry, one needs to have an exterior that can develop an
event horizon. Despite, we did not check the apparition of the event horizon, one can rely on the Penrose cosmic censorship hypothesis which states that the generic singularities arising in the gravitational collapse of a physically reasonable matter are not naked and are always black hole singularities. In other words, every singularity must possess an event horizon that hides the singularity from view. If we rely on the Penrose cosmic censorship hypothesis, then the mass of the black hole is smaller compared to the one we get if we consider a continuum model of the EOS. This is of great interest for studying the formation of primordial black holes PBH (black holes that have formed just after the Big Bang) because PBHs with smaller masses are good candidates for providing dark matter \cite{Carr:2019,Carr/etal:2016}.

\begin{table}
	\centering
	\caption{limits of stability of self-gravitating clusters in $3+1$ dimensions at different temperatures.}
	\label{tab:stability_limits}
	\begin{tabular}{lccr} 
		\hline
		$\bar{T}$ & $M_T(10^{-3}\hspace{0.5mm} M_{\odot})$ & $R(R_0)$\\
		\hline
		0.1  & 1.34112 & 9.53034 \\
		0.33 & 2.83079 & 12.3493 \\
		1    & 3.62104 & 13.771\\
		5    & 2.19519 & 12.2084\\
		7    & 1.69156 & 11.2843\\
		\hline
	\end{tabular}
\end{table}
\subsection{Higher Dimensions}
\begin{figure}[h]
\centering
\includegraphics[scale=0.27]{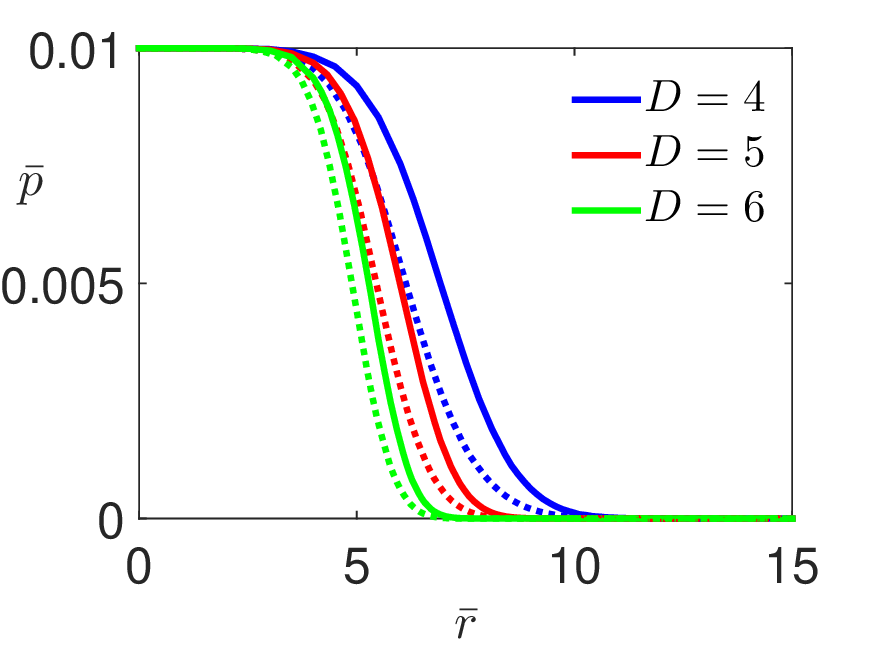}
\includegraphics[scale=0.27]{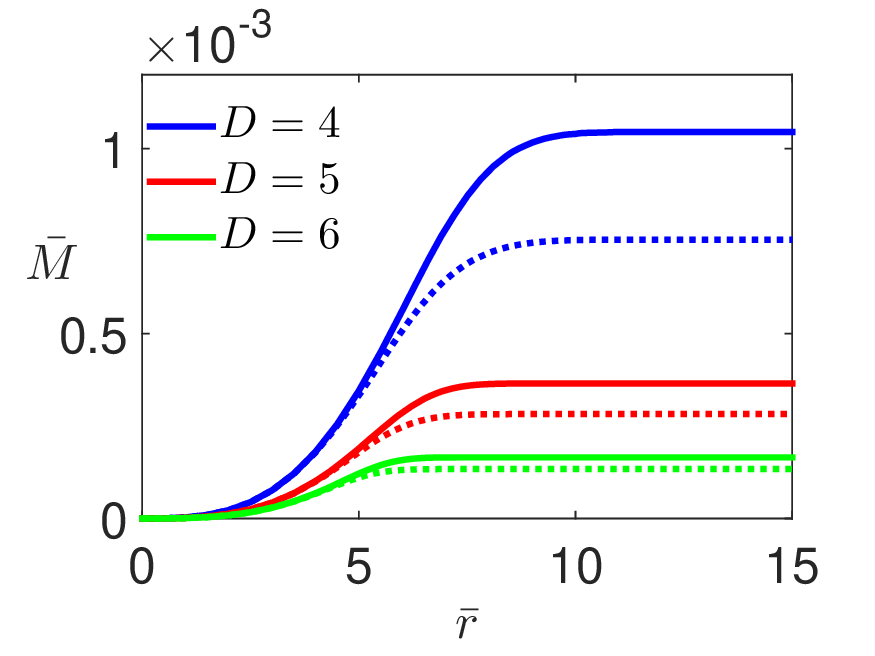}
\caption{\label{fig:fig4} Pressure $\bar{p}$ and mass $\bar{M}=M/M_\odot$ versus scaled distance $\bar{r}=r/R_0$ at different temperatures for self-gravitating lattice system in $D=4$, $5$ and $6$ dimensions. Solid curves represent the solution at $\bar{T}=1.0$. Dashed curves represent solutions at $\bar{T}=0.33$. Similar colors refer to the same dimensionality.}
\end{figure}

The numerical solutions of Eq.~(\ref{eq:d_tov_scaled}) in higher dimensions ($D=4$, $5$ and $6$) are shown in Fig.~(\ref{fig:fig4}). 
We see that both the mass and radius of the compact object decrease with increasing dimensionality. But as in three-dimensions, the gravitational attraction in higher dimensions leads to the formation of a black hole with a smaller mass and radius. In both panels, we have considered $\bar{T}=0.33$ (dashed lines) and $\bar{T}=1$ (solid lines). We see by comparing Fig.~(\ref{fig:fig2}) and (\ref{fig:fig4}), that as has been predicted by the polytropic EOS \cite{Leon/Cruz:2000}, the effect of gravity is stronger in dimension $D=3$ than any other dimension. The mass-radius curves in higher dimensions have qualitatively the same structure as the one in $3+1$ dimensions except that the compact objects have a smaller sizes. 
\section{Friedmann Equations}\label{sec:fr_eq}
To get the time evolution of the density $\rho$, pressure $p$ and the scale factor $a$ (that characterizes the expansion of the Universe), Eq.~(\ref{eq:eq_state}) needs to be combined with the Friedmann equations \cite{Friedman:1922}. The latter represent a general relativistic description of the expansion of space in homogeneous and isotropic models. In the present case, Eq.~(\ref{eq:metric}) reduces to the $D+1$-dimensional Friedmann-Lema\^itre-Robertson-Walker (FLRW) metric
\begin{align}\label{eq:flrw_metric}
ds^2=-dt^2+a^2(t)g_{ij}dx^{i}dx^{j},
\end{align}
where $g_{ij}$ is the metric of $D$-dimensional Riemannian manifold of constant scalar curvature. It is given by
\begin{align}\label{eq:flrw_spa_metric}
g_{ij}=\frac{1}{1-kr^2}dr^2 + r^2d\Omega_{D-1}.
\end{align}
The curvature of the Universe is characterized by the constant $k$ which
takes only three values depending on the curvature. When
$k = +1$, the Universe has a positive curvature and finite size,
while for $k = 0$, it is Euclidean flat and infinite. For $k = -1$,
the spacetime is infinite but has a hyperbolic geometry. Inserting the metric Eq.~(\ref{eq:flrw_metric}) and the stress-energy tensor Eq.~(\ref{eq:stress_energy_tensor}) into the $D$-dimensional Einstein field equations
\begin{align}\label{eq:einstein_1}
G_{\mu\nu}+\Lambda g_{\mu\nu}=8\pi G_DT_{\mu\nu},
\end{align}
and into the equation of conservation for the stress-energy tensor ($\nabla_{\mu}T^{\mu\nu}=0$), we get the Friedmann equations. For a homogeneous and isotropic Universe in $D+1$ dimensions, they are given by \cite{Chen/etal:2014}
\begin{align}\label{eq:friedman_eq}
&\dot{\rho}+D\frac{\dot{a}}{a}\left(\rho + \frac{p}{c^2}\right)=0,\\
&\dot{H}=-\frac{8\pi G_D}{D-1}\left(\rho+\frac{p}{c^2}\right)+\frac{kc^2}{a^2},\\
&H^2=\frac{16\pi G_D}{D(D-1)}\rho - \frac{kc^2}{a^2}+\frac{2\Lambda c^2}{D(D-1)},
\end{align}
in which the dot denotes time derivative and
\begin{align}
H=\frac{\dot{a}}{a}.
\end{align}
is the Hubble constant. The parameter $\Lambda$ is the Einstein cosmological constant. Let us scale all quantities in terms of the Planck units, 
\begin{align}
\bar{\rho}=\frac{\rho}{\rho_P},\hspace{5mm} \bar{a}=\frac{a}{l_P},\hspace{5mm}\bar{t}=\frac{t}{t_P},\hspace{5mm}\bar{p}=\frac{p}{\rho_Pc^2},
\end{align}
where $\rho_P=5.16\hspace{0.5mm}10^{99}g/m^3$ is the Planck density, $l_P=1.62\hspace{0.5mm}10^{-35}$ is Planck length and $t_P=5.39\hspace{0.5mm}10^{-44}$ is the Planck time. 
With this new scaling, the Friedmann equations (\ref{eq:friedman_eq}) become
\begin{align}\label{eq:friedman_eq_scal}
&\dot{\bar{\rho}}+D\frac{\dot{\bar{a}}}{\bar{a}}\left(\bar{\rho} + \bar{p}\right)=0,\\
&\dot{H}=-\beta_D\left(\bar{\rho}+\bar{p}\right)+\frac{\bar{k}}{\bar{a}^2},\\
&H^2=\frac{1}{D}\beta_D\bar{\rho} - \frac{\bar{k}}{\bar{a}^2},
\end{align}
where 
\begin{align}
\beta_D=\frac{4S_D R_0c^2\rho_P}{(D-1)M_{\odot}},\hspace{10mm}\bar{k}=\frac{kc^2}{l_P^2},
\end{align}
and we have considered the case $\Lambda=0$. The Hubble constant becomes
\begin{align}
H=\frac{1}{t_P}\frac{1}{\bar{a}}\frac{d\bar{a}}{d\bar{t}}=\frac{1}{t_P}\frac{\dot{\bar{a}}}{\bar{a}}.
\end{align}
Numerical solutions require for Eq.~(\ref{eq:friedman_eq_scal}) to be supplemented by closer relation which is the EOS.
\begin{figure}[htb]
\centering
\includegraphics[scale=0.4]{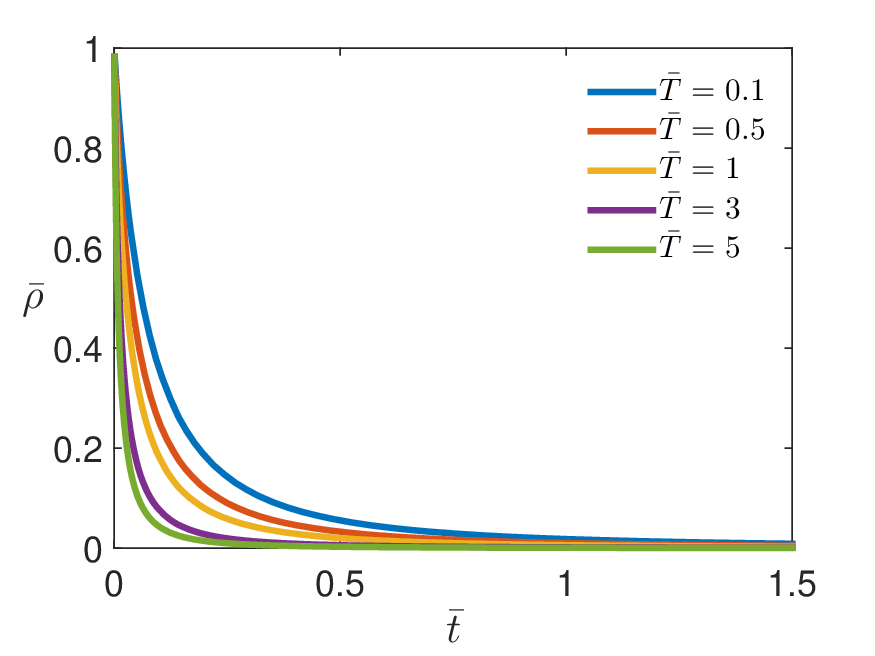}
\caption{\label{fig:fig5} Scaled density $\bar{\rho}$ versus scaled time at different temperatures of a lattice system in $2+1$ dimensions with $k=0$.}
\end{figure}
\subsection{2+1 Dimensions}
\begin{figure}[htb]
\centering
\includegraphics[scale=0.27]{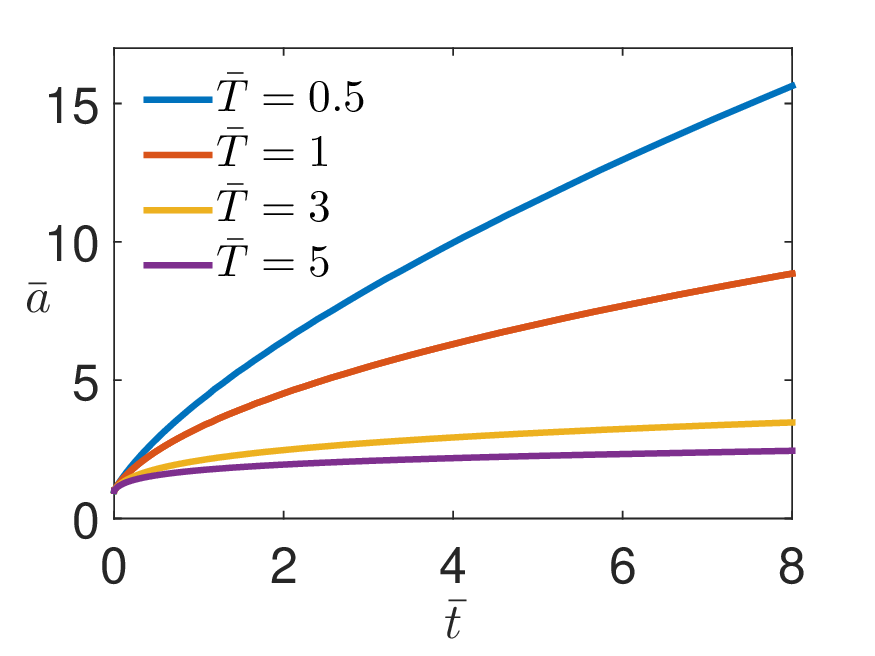}
\includegraphics[scale=0.27]{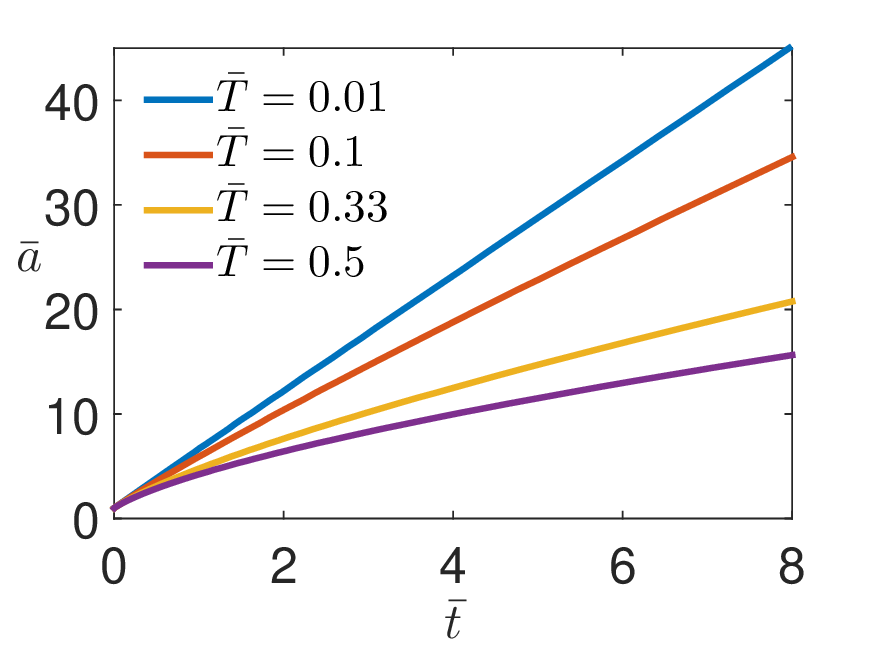}
\caption{\label{fig:fig6} Scale factor $\bar{a}$ versus scaled time $\bar{t}$ at high (left) and low (right) temperature of a lattice system in $2+1$ dimensions with $k=0$.}
\end{figure}

\begin{figure}[htb]
\centering
\includegraphics[scale=0.27]{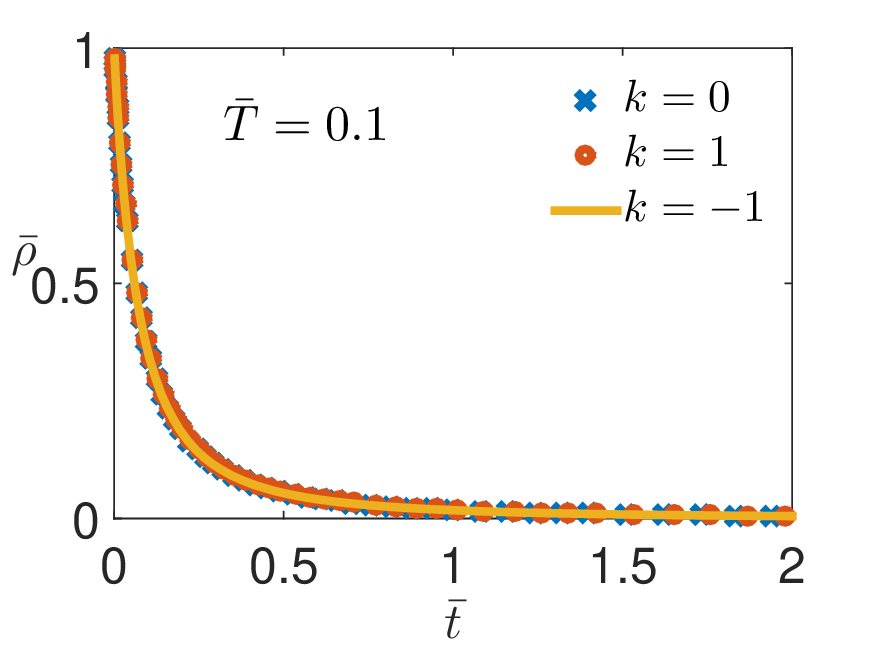}
\includegraphics[scale=0.27]{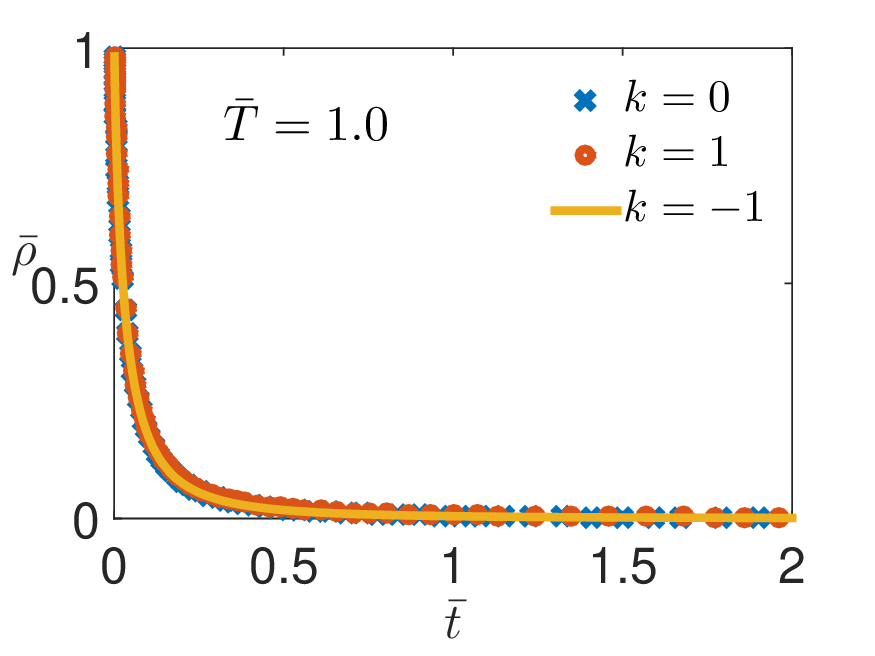}\\
\includegraphics[scale=0.27]{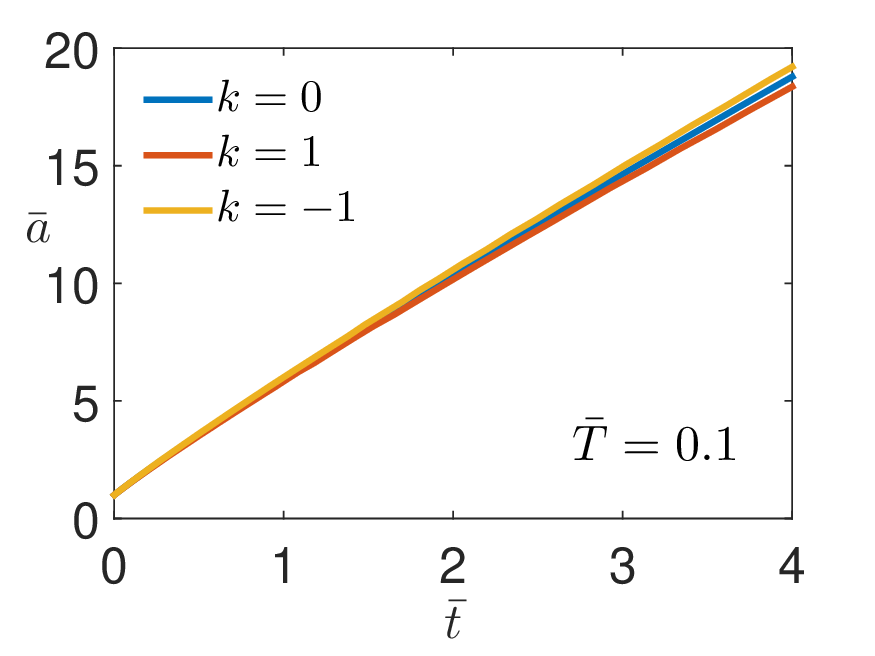}
\includegraphics[scale=0.27]{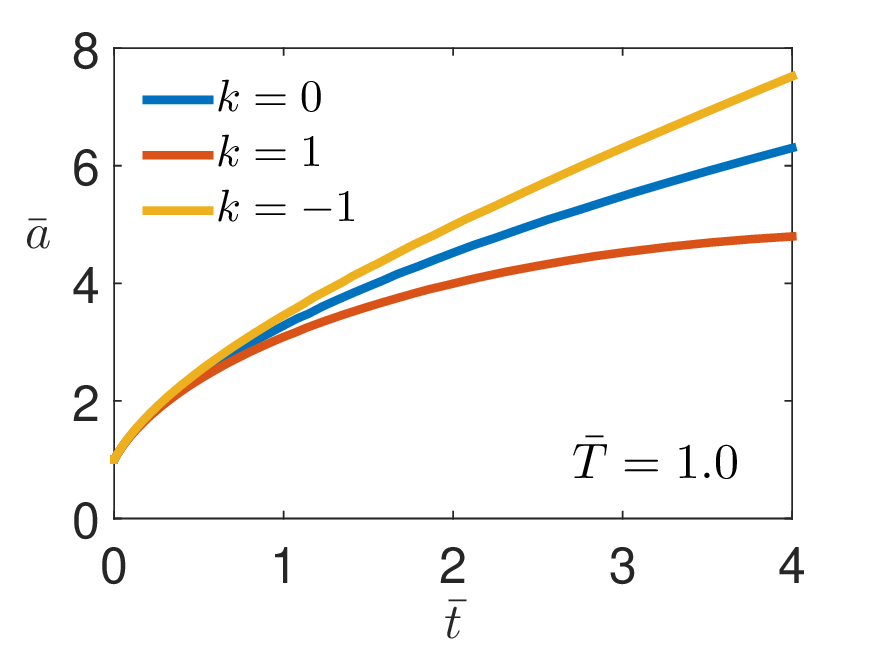}
\caption{\label{fig:fig7} Density and Scale factor versus scaled time at high and low temperature for $2+1$ dimensional lattice system with different curvatures.}
\end{figure}

In $2+1$ dimensions, the Friedmann equations become
\begin{align}\label{eq:friedman_eq_2d}
&\dot{\bar{\rho}}+2\frac{\dot{\bar{a}}}{\bar{a}}\left(\bar{\rho} + \bar{p}\right)=0,\\
&\dot{H}=-\beta_2\left(\bar{\rho}+\bar{p}\right)+\frac{\bar{k}}{\bar{a}^2},\\
&H^2=\beta_2\bar{\rho}-\frac{\bar{k}}{\bar{a}^2}.
\end{align}
The numerical solutions of Eqs.~(\ref{eq:friedman_eq_2d}) for the case of flat space $k=\bar{k}=0$ (Einstein-de Sitter Universe) are shown in Fig.~(\ref{fig:fig5}) and (\ref{fig:fig6}). As shown in Fig.~(\ref{fig:fig5}), the scaled density drops rapidly with time due to the expansion of the Universe. The scale factor increases at a decreasing rate. In this case, the density is equal to a critical value at which the Universe will expand forever at a decreasing rate.
Variations of the density profiles and scale factor as a function of cosmological time at high and low temperatures are shown in Fig.~(\ref{fig:fig7}) for different curvatures of the spacetime. Time evolution of density profile has almost a universal form with respect to space curvature and temperature. However, the scale factor depends strongly on the combined effects of temperature and curvature. At low temperatures (including the case of $\bar{T}=0.33$ corresponds to the lattice version of the radiation-dominated era), the curvature has almost no effect on the scale factor, meaning that the Universe expands in the same way. But at high temperatures (right panel in Fig.~(\ref{fig:fig7})), the Universe expands faster when the curvature increases. For both a flat space and curved space with negative curvature, the Universe expands to infinity at a decreasing rate. For positive curvature, the Universe has a finite size and this has been shown also in the continuous model for the EOS.
\subsection{3+1 Dimensions}
\begin{figure}[htb]
\centering
\includegraphics[scale=0.4]{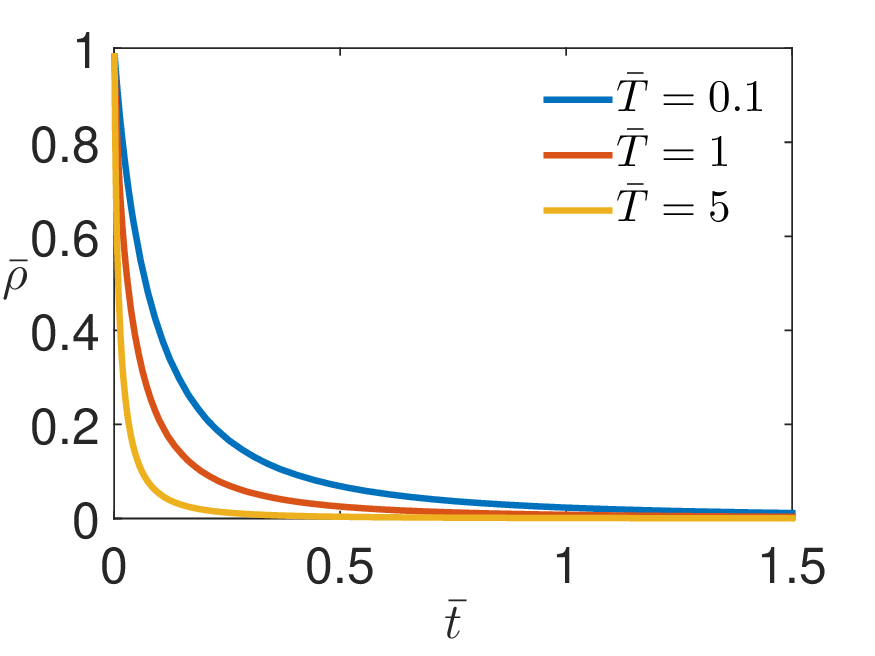}
\caption{\label{fig:fig8} Scaled density $\bar{\rho}$ versus scaled time for different temperatures of self-gravitating gas in $3+1$ dimensions.}
\end{figure}

\begin{figure}[htb]
\centering
\includegraphics[scale=0.27]{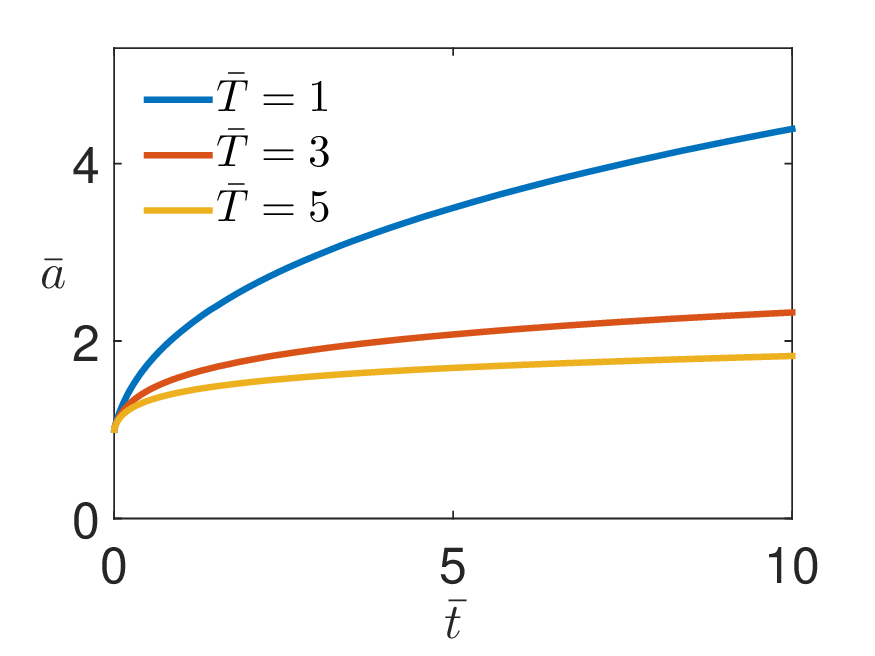}
\includegraphics[scale=0.27]{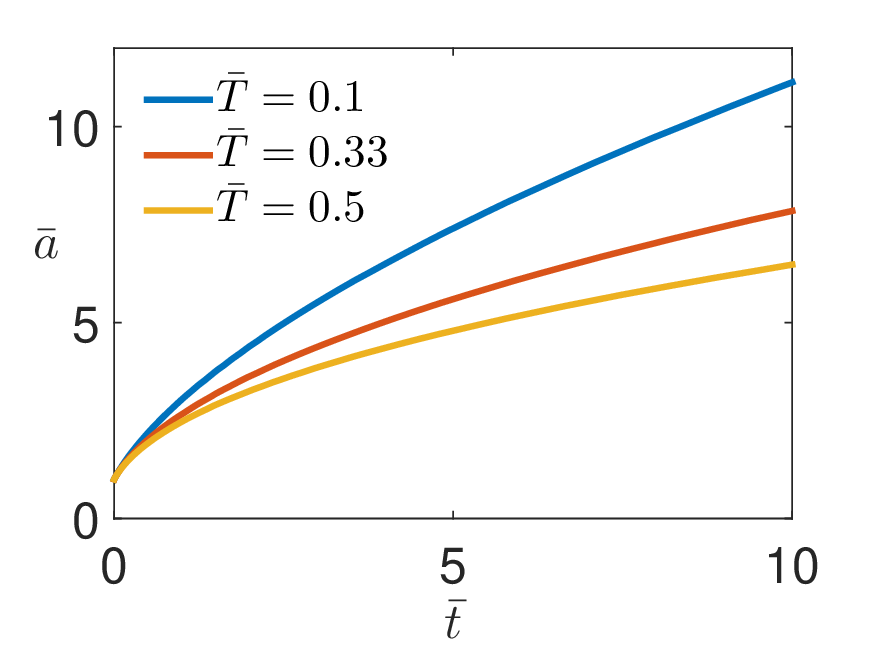}
\caption{\label{fig:fig9} Scale factor $\bar{a}$ versus scaled time $\bar{t}$ at high (up) and low (down) temperature of self-gravitating gas in $3+1$ dimensions.}
\end{figure}

\begin{figure}[htb]
\centering
\includegraphics[scale=0.27]{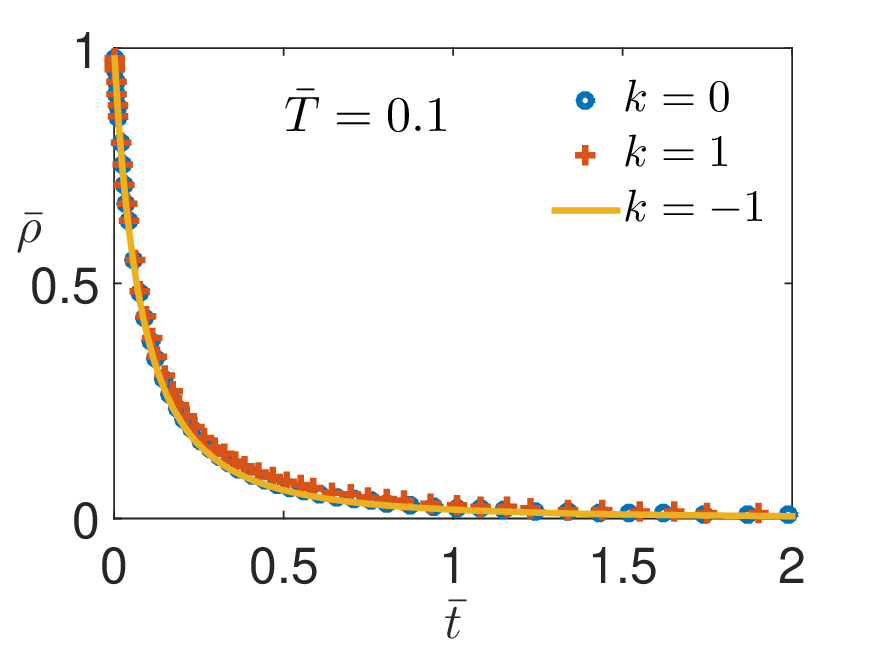}
\includegraphics[scale=0.27]{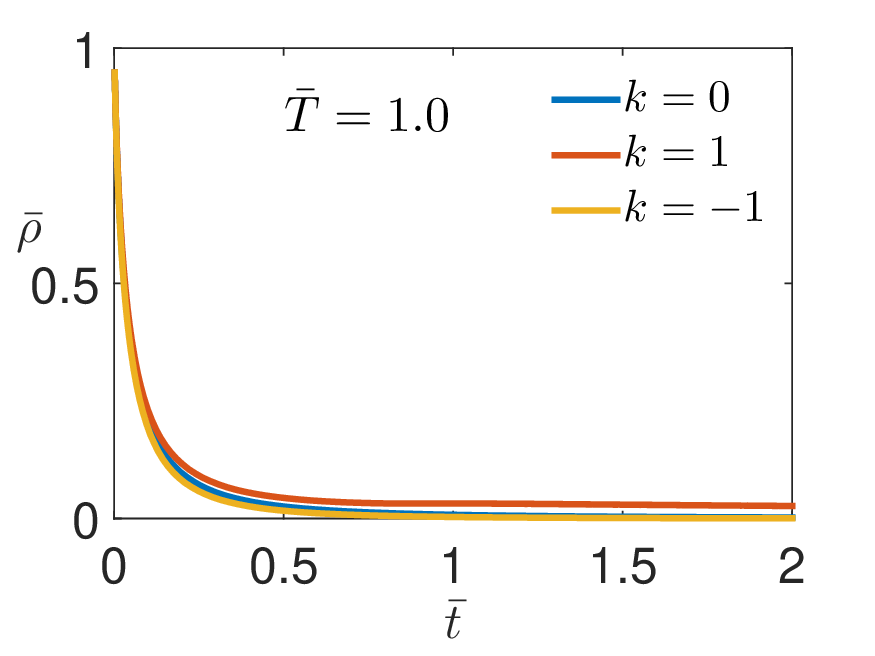}\\
\includegraphics[scale=0.27]{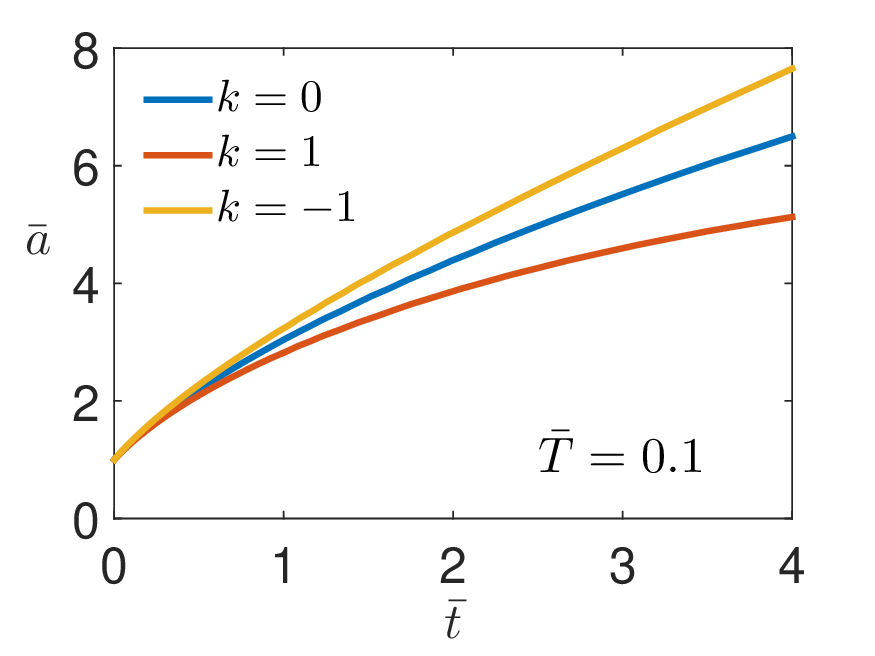}
\includegraphics[scale=0.27]{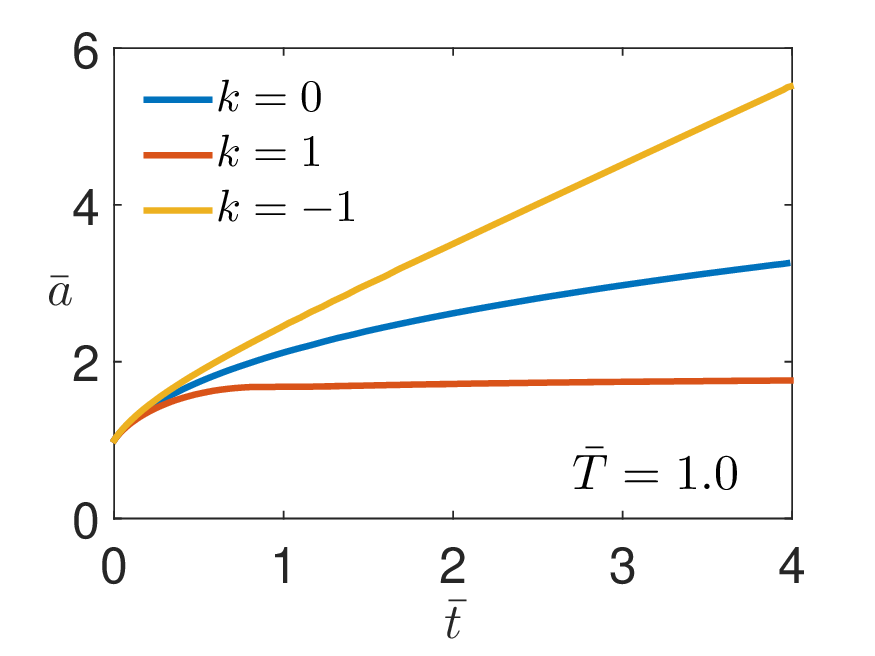}
\caption{\label{fig:fig10} Density and Scale factor versus scaled time at high and low temperature for $3+1$ dimensional spacetime with different curvatures.}
\end{figure}

In spacetime of $3+1$ dimensions, the Friedmann equations are reduced to
\begin{align}
&\dot{\bar{\rho}}+3\frac{\dot{\bar{a}}}{\bar{a}}\left(\bar{\rho} + \bar{p}\right)=0,\\
&\dot{H}=-\beta_3\left(\bar{\rho}+\bar{p}\right)+\frac{\bar{k}}{\bar{a}^2},\\
&H^2=\frac{1}{3}\beta_3\bar{\rho}-\frac{\bar{k}}{\bar{a}^2}.
\end{align}
and their numerical solutions are shown in Fig.~(\ref{fig:fig8}) and Fig.~(\ref{fig:fig9}). The two figures show that profiles for mass density and scale factor are similar to the case of $D=2$, but quantitatively they are different. Fig.~(\ref{fig:fig10}) shows that the scale factor varies with time as $\bar{a}\sim \bar{t}^{1/2}$ for $\bar{T}=0.33$ and as $\bar{a}\sim \bar{t}^{2/3}$ for $\bar{T}=1$. This confirms the results predicted by the polytropic EOS.

\subsection{Higher Dimensions}
\begin{figure}[htb]
\centering
\includegraphics[scale=0.27]{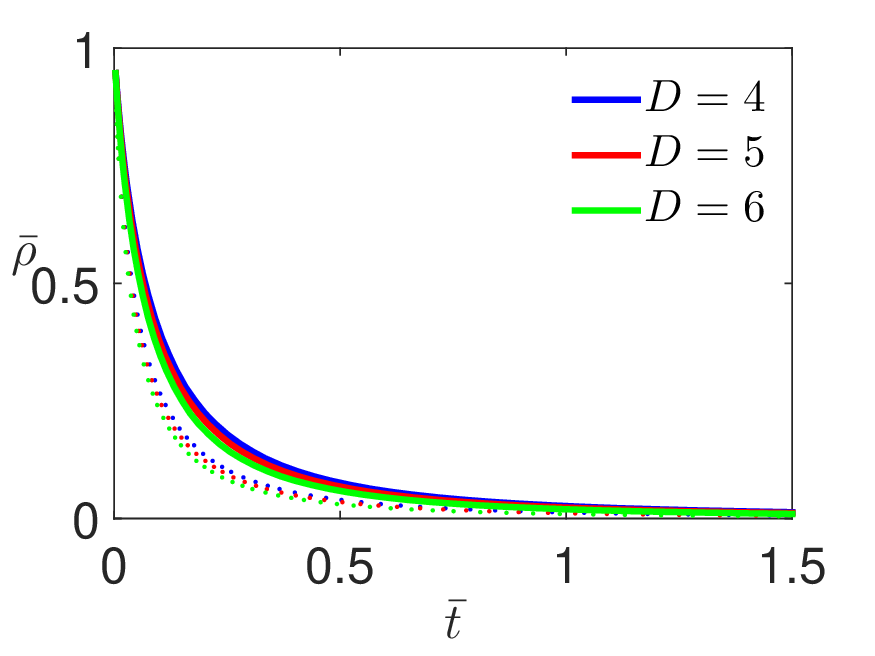}
\includegraphics[scale=0.27]{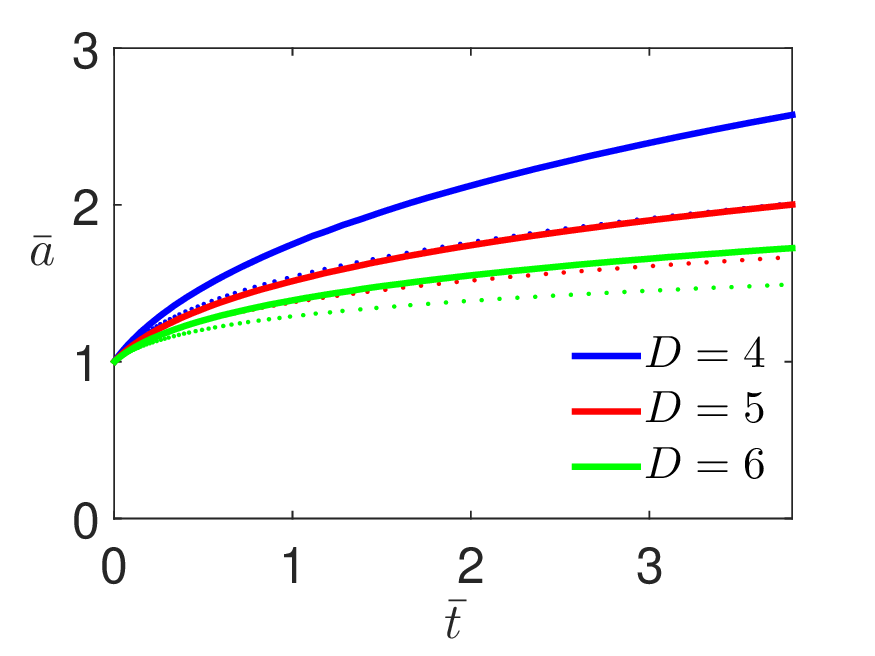}
\caption{\label{fig:fig11} Density profile $\bar{\rho}$ and scale factor $\bar{a}$ versus scaled time $\bar{t}$ at different temperatures for self-gravitating lattice system in $D=4$, $5$ and $6$ dimensions. Solid curves represent the solution at $\bar{T}=0.33$. Dashed curves represent solutions at $\bar{T}=1.0$. Similar colors refer to the same dimensionality.}
\end{figure}

Numerical solutions of the Friedmann equations in dimensions $D = 4$, $5$ and $6$ are shown in Fig.~(\ref{fig:fig11}). As can be seen, dimensionality has week effect on both density and scale factor at either $\bar{T}=0.33$ corresponds to the "radiation-dominated era" and $\bar{T}=1.0$.
\section{Conclusion}\label{sec:conc}
In this work, we have investigated the possibility of the formation of stellar compact objects using a lattice EOS, and the results are compared to those of the polytropic continuum EOS that has been used in previous studies. The ratio $M/R$ of the compact object found here falls within the range of the one reported in the literature \cite{Silbar/Reddy:2003}. However, our results for the mass $M$ and radius $R$ are different from the ones that have been found using the standard calculation of GR, because of the effect of exclusion introduced by the lattice. At low-density, our analysis reduced neatly to all the results derived previously \cite{Chavanis:2014b,Bakhti/etal:2018}. At this limit, the lattice EOS is reduced to those used in the description of relativistic Neutron stars as well as the three eras of the early Universe (The inflation era, the radiation-dominated era, and the matter-dominated era). Numerical solutions of the TOV equation suggest that the mass and radius of the compact object depend strongly on the central pressure and it is generally smaller compared to the one found using a polytropic EOS. However, in $D=2$, the mass is independent of the central pressure. Hence only compact objects with finite and constant masses do exist and there is no possibility for the formation of black holes. In $D\geq 3$, there is a possibility for the formation of compact objects with a large mass gap. Our calculations show that gravity has a stronger effect in $3+1$ dimensions than any other dimension.
Beyond a certain mass threshold, the compact object becomes unstable and collapses into a singularity. This could be a sign of the formation of a black hole. The smaller mass weakens the gravitational attraction and hence the radius is greater compared to the continuous description. Calculations based on string theory and quantum gravity point to the existence of a black hole in $2+1$ dimensional anti-de Sitter space. Thus, it is of great interest to see how does the lattice gas EOS work with models of string theory and quantum gravity. We expect that models  based on  string theory and quantum gravity could  give some corrections to our model for stellar black hole. As a continuation  to this work, we are planning to use these advanced models combined with the lattice EOS to study the PBH where the calculation based on the TOV equation is not possible. We found that the lattice EOS imposes a lower limit on the mass of the stellar black hole and this could be very promising in studying the evaporation of the PBH and its connection to dark matter.

\bibliographystyle{apsrev4-1}
\bibliography{mybib}
\nocite{*}

\end{document}